\documentclass[runningheads]{cl2emult}

\usepackage{makeidx}  
\usepackage{graphicx} 
\usepackage{subeqnar} 
\usepackage{multicol} 
\usepackage{cropmark} 
\usepackage{eso}      
\makeindex            

\newcommand{\lta} {\mathrel{\hbox{\rlap{\hbox{\lower4pt\hbox{$\sim$}}}\hbox{$<$}}}}
\newcommand{\gta} {\mathrel{\hbox{\rlap{\hbox{\lower4pt\hbox{$\sim$}}}\hbox{$>$}}}}
\newcommand{\fesc} {f_{\rm esc}}
\newcommand{\Frat} {F_{9/15}}

\begin{document}
\title*{The Hubble Deep Field in the Far Ultraviolet}
\toctitle{The HDF in the Far Ultraviolet}
%
%
\titlerunning{The HDF in the Far Ultraviolet}
%
\author{Henry C. Ferguson}
\authorrunning{Henry C. Ferguson}
%
%
\institute{Space Telescope Science Institute \\
3700 San Martin Drive\\
Baltimore, MD 21218 USA
}

\maketitle              

\begin{abstract}
Results from a recent HST survey of field galaxies at wavelengths 1600\,{\AA} 
and 2400\,{\AA} are be presented. The data are used to constrain the
fraction of Lyman-continuum radiation that escapes from galaxies
at redshifts $z \approx 1$. The combined UV-IR photometry for
HDF galaxies is also used to investigate whether
low-mass starburst galaxies dominate the field-galaxy population
at redshift $z \approx 1$. The relative lack of objects with the
colors of faded bursts suggests that star-formation
is largly quiescent rather than bursty or episodic.

\end{abstract}

When the ultimate self-consistent model of cosmology and galaxy evolution
comes along it will be able to reproduce the full distribution of 
galaxy shapes, sizes, masses, and spectral types, at all redshifts and
in all environments. Until that happens, we as observers are
forced to try to piece together portions of the puzzle, gathering data
as new facilities and techniques become available, and testing and
refining parts of the theory. One of the new facilities that has relatively
recently become available is the Space Telescope Imaging Spectrograph
(STIS), which was installed on the Hubble Space Telescope in 
1997.  Prior to STIS, the deepest observations of galaxies at far-UV
($\lambda \lta 2500$\,{\AA})
wavelengths reached to $m \sim 20.7.$ 
\footnote{Magnitudes throughout this contribution are on the AB system
(Oke 1974), where $AB = -2.5 \log f_\nu ({\rm nJy}) + 31.4$} Far-UV
observations provide a sensitive probe of the population of young
massive stars in galaxies, and spectral breaks at Lyman $\alpha$ and
the Lyman break (912\,{\AA}) are among the strongest features that can be
used for estimating photometric redshifts. The STIS
Multi-Anode-Microchannel Array (MAMA) detectors provide sensitive
imaging over a $25^{\prime\prime} \times 25^{\prime\prime}$ in two
bands centered at approximately 1600\,{\AA} (FUV) and 2400\,{\AA} (NUV).
Rest-frame emission shortward of 2000\,{\AA} can be measured for the
first time in galaxies in the crucial redshift range $0.5 < z < 2.5$,
probing the period where most of the stars in the present-day 
universe were formed.

Amongst the aims of our survey were to (1) measure galaxy counts in the
far-UV, (2) compare rest-frame UV morphologies to optical morphologies
for galaxies at $0.5 \lta z \lta 2$, (3) measure the far-UV background
and constrain the contribution from faint galaxies, (4) constrain
the contribution of galaxies to the metagalactic ionizing background
radiation, and (5) constrain the star-formation histories of dwarf
galaxies at redshifts $z \lta 1$. 
The observations cover an
area of one square arcminute, with three sigma detection limits of
$AB = 29.1$ in both the NUV and FUV bands. 
Galaxy counts from our survey have
been presented by Gardner et al. \cite{GBF00} and constraints on the
FUV background by Brown et al. \cite{BKFGCH00}.
In this contribution, we concentrate on two topics: setting
constraints on the ionizing radiation escaping from galaxies and
setting constraints on star-formation histories of low-luminosity
galaxies in the HDF.

\section{The Ionizing Radiation Escaping from Galaxies}

Studies of quasar absorption lines indicate that the intergalactic
medium is highly ionized out to redshifts of order $z=5$.
Instead of appearing as a smooth Gunn-Peterson (1965) absorption
trough at redshifts less than 1216\,{\AA} in the rest frame QSO of
the QSOs, the absorption due to neutral hydrogen in the IGM appears
at discrete wavelengths (the Lyman-alpha forest). The total baryon
content in neutral gas is much less than that expected
from primordial nucleosynthesis, and both the measurements of the line
strengths of different ionization states heavier elements,
and the measurement of the proximity effect suggest that the 
gas is highly ionized by a metagalactic radiation field 
$J_\nu \sim 10^{-21}\rm \,erg\,s^{-1}\,cm^{-2}\,Hz^{-1}\,sr^{-1}$
\cite{SS89,BDO88}. It is likely that the ionization
of the IGM plays an important role in regulating the cooling and
collapse of gas to form galaxies \cite{BR92,TW96}.

The IGM at high redshifts is most likely ionized by young galaxies and
QSOs. However, the relative contribution of these two types of objects
to the high-redshift ionizing background is not well constrained.  The
declining space density of QSOs beyond $z=3$ suggests that even if they
produce most of the ionizing photons at $z \sim 2$, their contribution
is likely to decline dramatically at high redshifts. An extrapolation
of the measured luminosity functions and redshift evolution 
suggests QSOs and other AGN are likely to provide only about 
25\% of the ionizing photons at 1 Ryd at z=5.  In
contrast, the Lyman continuum flux estimated to arise from the
integrated LF of galaxies at $z \sim 3$ appears sufficient to keep the
IGM ionized, provided that a fraction $\fesc > 50\%$ of the ionizing
photons escape \cite{MHR99}.

The most direct constraints on $\fesc$ at redshift $z=0$ come from observations
of four starburst galaxies with the Hopkins Ultraviolet Telescope (HUT)
\cite{LFHL95}. 
More recently Steidel et al. \cite{SPA00} have reported a detection of Lyman
continuum radiation from a composite spectrum of 29 Lyman-break galaxies
at redshifts $z = 3.40 \pm 0.09$. The measured ratio of 900\,{\AA} to
1500\,{\AA} emission $\Frat$ is remarkably high, indeed higher than
expected from some population synthesis models even if 100\%
of the of the intrinsic Lyman continuum radiation escapes.  However, the
measurement of this break in the composite spectrum is extraordinarily
difficult, and a substantial correction for IGM attenuation is required
to derive the intrinsic flux decrement of the composite galaxy spectrum.
Even if the detection itself is real, there are concerns that the sample
is biased toward the bluest of the Lyman-break galaxies.

There are seven galaxies in our STIS HDF field with spectroscopic
redshifts $1<z<1.3$. At $z=1.1$ the center of the FUV bandpass
is at $\lambda_{\rm rest} \sim 760$\,{\AA}, and the integrated
throughput from wavelengths longer than  $\lambda_{\rm rest} = 912$\,{\AA} is 
less than 0.2\% of the total throughput.
To maximize the signal-to-noise ratio, 
UV fluxes and upper limits for the seven galaxies have been estimated using 
a $\chi^2$ fitting technique (TFIT) 
\cite{PD00} on the STIS images after convolution to
the resolution of the F450W HDF images. 

\begin{figure}
\centering
\includegraphics[width=1.0\textwidth]{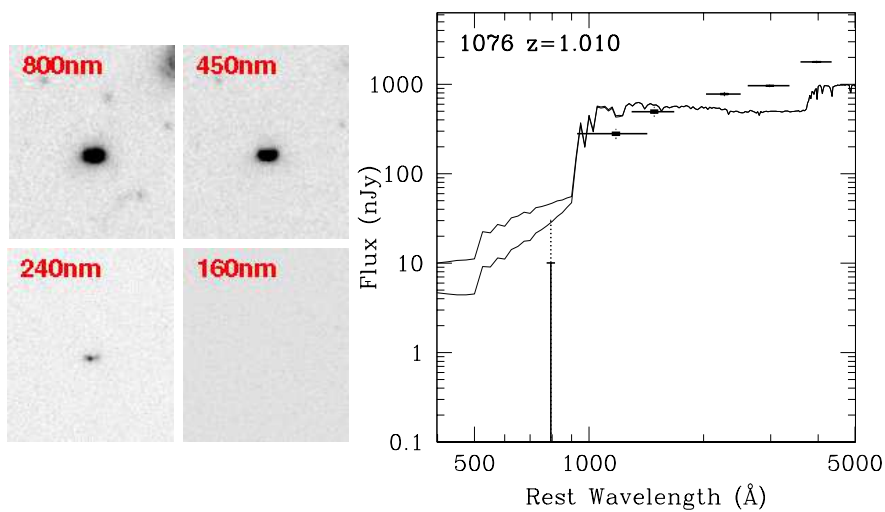}
\caption[]{Photometry for an HDF galaxy at $z=1.01$. The points with
errorbars show the measured photometry. The galaxy is undetected
in the shortest wavelength point; the 1$\sigma$ uncertainty
is shown as a solid error bar, and the 2$\sigma$ uncertainty as a 
dotted error bar.
The curves show a fiducial
model SED (not fit to the data) for an unobscured galaxy of age
$10^9$\,yr undergoing constant star formation with solar metallicity,
and a Salpeter IMF. The top curve is the emergent spectrum from
the galaxy; the bottom curve is attenuated by the IGM.
}
\label{eps1}
\end{figure}

Figure 1 shows the photometry for one of the galaxies, superimposed on
a typical star-forming galaxy spectral energy distribution with
and without Madau \cite{Madau95} IGM attenuation. Only one of
the seven galaxies in our sample is detected, and that one 
at only the 2.5$\sigma$ level.

For a galaxy at $z=1$, the half-maximum points of the FUV bandpass
are at rest-frame wavelengths 730 and 830\,{\AA}, hence a small
correction is necessary to estimate F(900). We have so far done this
only in a crude way, accounting for the typical IGM attenuation
but assuming a flat spectrum in $f_\nu$ rather than a model
stellar population. The resulting values of $\Frat$, displayed in
Figure 2, should thus be regarded as preliminary, but are unlikely
to change by more than a few percent.

\begin{figure}
\centering
\includegraphics[width=0.6\textwidth]{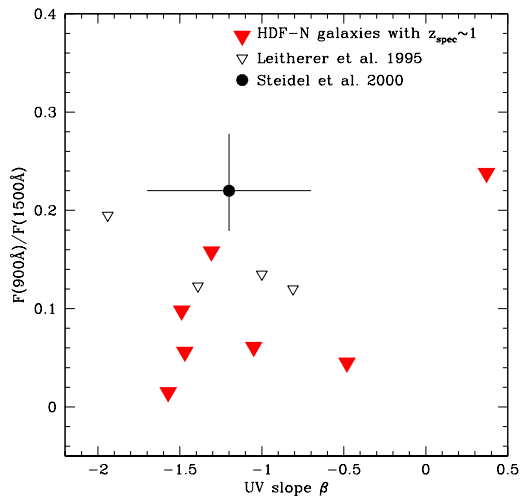}
\caption[]{Ratio $\Frat$ vs. UV spectral slope 
$\beta \,\,(f_\lambda \propto \lambda^\beta)$, for the HDF 
galaxies (solid triangles) compared to the results from Leitherer 
et al. \cite{LFHL95} (open triangles) and Steidel et al. \cite{SPA00}
(solid circle). 
The value of $\beta$ has not been measured for the Steidel et al.
composite galaxy, and it has been placed arbitrarily at $\beta = 1.2$.
For the HDF sample $\beta$ is estimated from the photometry at
$\lambda_{\rm rest} < 2500$\,{\AA}, while for the Leitherer et al. 
sample it is measured from the HUT spectra in absorption-line free
regions with $1220 < \lambda < 1800$. Nearly all of the upper
limits for galaxies observed with HST or HUT fall below the 
Steidel et al. \cite{SPA00} measurement.

}
\label{eps2}
\end{figure}

For unreddened star-forming objects, population synthesis models
predict $\Frat \approx 0.2$, which is roughly the minimum level
of Lyman continuum jump consistent with the Leitherer et al. \cite{LFHL95}
spectra.  Leitherer et al. used H$\alpha$ measurements of the 
galaxies in their sample to set a lower limit on the total number of 
Lyman-continuum photons that must be present. Based on that analysis
they derived upper limts on $\fesc$ ranging from 1\% to 15\%. Lacking
H$\alpha$ fluxes for the HDF galaxies, we must adopt a model spectrum
to infer $\fesc$ from $\Frat$. For this preliminary analysis we adopt
the model shown in Fig. 1; in the final analysis this will be refined
by fitting models to each individual galaxy. For this model,
if we assume that all the UV radiation produced by stars at rest-frame
1500\,{\AA} escapes from the program galaxies, then our upper limits
translate into $\fesc < 0.2$ to 0.9 (depending on the galaxy). If we
instead make the more reasonable assumption that there is some
dust attenuation of the stellar radiation at both 900\,{\AA} and
1500\,{\AA}, then adopting a factor of 4.7 attenuation at 1500\,{\AA}
\cite{SAGDP99}
and a Calzetti \cite{Calzetti97} attenuation curve leads to upper limits
$\fesc < 0.05$ to $0.2$, significantly below the claimed detection
\cite{SPA00}, but compatible with the expectations from recent radiative
transfer models \cite{DSF00}. In general, accounting for dust and
gas attenuation, $\fesc < 20$\,\% for all galaxies with sufficiently
deep measurements from both HUT and HST. 

\section{Star-formation in galaxies at $z \sim 1$}

It is likely that the ionizing background inhibits the cooling of
gas in low-density regions, and in particular in the halos of
low-mass galaxies \cite{Efstathiou92,BR92,TW96,NS97}. 
This has led to suggestions that a large fraction of the 
faint-blue galaxies are dwarf galaxies undergoing starbursts 
at moderate redshifts. While the most extreme of these models
are ruled out \cite{FB98}, it is still possible that dwarf
galaxies undergoing some form of episodic star formation are
a major contributor to faint-galaxy counts \cite{Campos97}.

\begin{figure}
\centering
\includegraphics[width=1.0\textwidth]{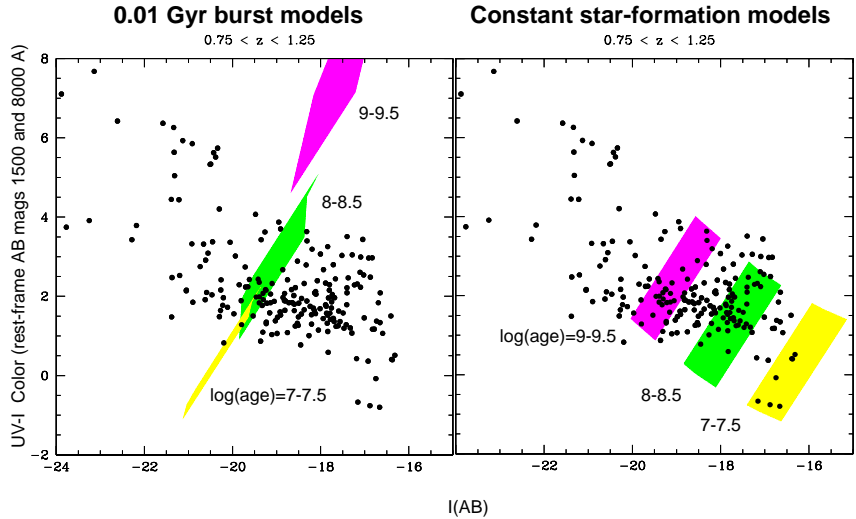}
\caption[]{Rest-frame $M$(1500{\AA})$ - I_{AB}$ color--magnitude
diagram for HDF galaxies with photometric redshifts $0.75 < z < 1.25$.
Galaxies are from an H-band selected catalog covering
the entire 5 square arcminutes of the HDF-N \cite{Detal2001}, and 
rest-frame magnitudes have been interpolated from the multi-band
photometry. Photometric redshifts were determined via template
fitting. Superimposed on the data are shown the areas that
should be occupied by galaxies in three ranges of 0.5 dex in age,
with Calzetti \cite{Calzetti97} extinctions $0.02 < E(B-V) < 0.4$.
The left panel shows starburst models. The stellar mass in the models
is $10^9 M_\odot$, the metallicity is 1/5 solar, and the star-formation
e-folding timescale is $\tau = 10^7$\,yr, which is roughly the timescale
expected for supernova winds to drive the gas from a dwarf galaxy. 
The right panel shows models with a constant star-formation rate
of $1 M_\odot\, {\rm yr^{-1}}$. Thus at an age of $10^9\,$yr the models
in both panels have the same mass. The basic point is that distribution
of the HDF galaxies in this color-magnitude plane makes sense if
their star formation is relatively constant, while it is hard to 
understand if their star formation is bursty or episodic.
}
\label{eps3}
\end{figure}

From the combined multi-wavelength HDF observations, it is possible
to investigate whether starbursts or episodic star formation are
ubiquitous or rare in the faint-galaxy population. The argument
goes as follows: if dwarf galaxies are forming stars in bursts
with star-formation timescales $\tau_{SF}$ and with intervals between
episodes $\Delta t_{\rm burst} >> \tau_{SF}$, then provided this is
roughly a steady-state process there ought to be 10 times as many
galaxies with ages $t \sim 10 \tau_{SF}$ as there are galaxies
with age $\tau_{SF}$. And there ought to be 100 times as many
galaxies with ages $t \sim 100 \tau_{SF}$ 
provided $\Delta t_{\rm burst}$ is greater than $t$.
We can thus examine the ``fading sequences'' in color-magnitude
or color-color space and see whether the galaxies are piling up
in location expected if they are bursting and then fading.

Figure 3 shows a color--absolute-magnitude diagram of HDF galaxies with 
photometric redshifts $0.75 < z < 1.25$, spanning a range of
lookback times from 7 to 9.2 Gyr for a cosmology with
$h, \Omega_{\rm tot}, \Omega_{\rm m}, \Omega_\Lambda = 0.65,1,0.3,0.7.$
The two panels of the figure show the regions that should be
occupied by galaxies with stellar masses of $10^9 M_\odot$ for
with Calzetti \cite{Calzetti97} extinctions $0.02 < E_{B-V} < 0.4$ 
and ages in three ranges of 0.5 dex. The left panel shows 
$\tau = 10^7\,{\rm yr}$ burst models and the right panel shows
constant star-formation models. Now, given that the number of
years spanned by the three regions increases by a factor of
10 as they move to older ages, any steady state process should
have the galaxies bunching up at the older ages. The distribution
of colors for the HDF galaxies is clearly {\it inconsistent}
with most of them undergoing short bursts. Under these
model assumptions the HDF galaxy colors cluster around ages of
$10^8\,{\rm yr}$, but there is no increase of galaxy numbers toward
older ages. In contrast, the loci of HDF points are reasonably
consistent with that expected from roughly constant star-formation
rates. From this type of analysis, a fairly robust conclusion 
is that {\it star-formation in faint-blue galaxies
at $z \sim 1$ is a relatively quiescent process. Dwarf galaxies
undergoing single bursts or episodic star formation with a duty
cycle $\Delta t \gta 10^9\,yr$ are not the dominant
population.} Alternatively, the bursting galaxies must fade
more rapidly than expected from simple population-synthesis models,
either because the IMF is top heavy, or because the stellar
population expands when the gas is driven out, and the galaxies then
drop below the HDF detection limit in surface brightness.

I would like to thank my collaborators on this project, particularly
Tom Brown, Jon Gardner, Eliot Malumuth, and Casey Papovich. 
Support for this work was provided by NASA through grant 
number GO-07410.01-96A from the Space Telescope Science Insitute,
which is operated by the Association of Universities for Research
in Astronomy under NASA contract NAS5-26555.

\clearpage
\addcontentsline{toc}{section}{Index}
\flushbottom
\printindex

\end{document}